# Slip of fluid molecules on solid surfaces by surface diffusion

## Jian-Jun Shu*, Ji Bin Melvin Teo, Weng Kong Chan

School of Mechanical & Aerospace Engineering, Nanyang Technological University, 50 Nanyang Avenue, Singapore 639798.

* Corresponding author

E-mail: mjjshu@ntu.edu.sg (JJS)

**Competing Interests:** The authors have declared that no competing interests exist.

**Financial Disclosure:** The authors received no specific funding for this work.

## Abstract

The mechanism of fluid slip on a solid surface has been linked to surface diffusion, by which mobile adsorbed fluid molecules perform hops between adsorption sites. However, slip velocity arising from this surface hopping mechanism has been estimated to be significantly lower than that observed experimentally. In this paper, we propose a re-adsorption mechanism for fluid slip. Slip velocity predictions *via* this mechanism show the improved agreement with experimental measurements.

## Introduction

The physical process of fluid slip on solid surfaces remains vague despite the plethora of experimental and theoretical studies [1-3]. A fairly clear picture of gas-solid slip can be derived within the kinetic theory framework [4]. In liquid-solid slip, however, the scattering model is inadequate due to the additional interactions with neighbouring liquid molecules from the bulk flow. At this stage, the contentious influences of surface nanobubbles and wetting, among other factors, have to be isolated before the primary mechanism(s) can be identified through experiments. Nevertheless, several plausible slip models have been put forward.

Tolstoi [5] was among the earliest to adopt a molecular kinetics approach for describing slip behaviour through the difference between surface and bulk liquid molecular mobilities, showing a link between slip and surface wettability. His work was later improved by Blake [6] to overcome its limitations in complete-wetting situations. The work of Ruckenstein & Rajora [7] was often quoted in the literature for their insightful suggestion that a surface gas layer could be a contributing factor towards the experimentally observed magnitudes of slip and otherwise could not be purely explained by their surface diffusion model. Yet, the attempt to the associate slip with the thermally activated motion of molecules on a substrate lattice deserved more plaudits. The slipping of the interfacial layer could be pictured as the surface diffusion with a net drift, comprising of a series of hops by the fluid molecules between substrate lattice sites while being subjected to an external field [8]. Lichter *et al*. [9] suggested a similar surface hopping mechanism in their rate theory model of slip flow. Their Arrhenius-type model was conceptually similar to the Blake-Tolstoi model but considered tilted potential barriers between the adsorption sites with lower barriers in the direction of the external field. This leaded to a net drift velocity that could be considered as the molecular slip velocity.

It was estimated that a very high shear rate of about $10^{12}$ s$^{-1}$ is required for slip to develop in this manner [1]. However, adopting the rate theory model where the hopping occurs by thermal vibration showed that it was not necessary for the hydrodynamic force to be greater than the dispersion forces for slip to occur. The slip velocity in the surface diffusion model appeared to show a non-linear dependence on slip but a rough estimate using appropriate parameters revealed that slip remained within the linear regime for the range of experimental shear rates [10]; the expression recovered the familiar Navier form. Slight adaptations to the model had also been made to include a critical shear stress criterion and shear-dependent dissipation at high shear rates to improve the match with results from numerical simulations but lacked strong physical justifications [11,12]. The one-dimensional Frenkel-Kontorova (FK) model had been used to represent the molecular mechanism of slip arising from the interplay of liquid-liquid and liquid-solid interactions [13,14]. The modified FK equation accounted for the mass flux in the direction normal to the surface, where the near-wall density was higher due to molecular ordering.

In this study, we explore an alternative surface diffusion process involving a re-adsorption mechanism that produces contrasting slip behaviour. The proposed molecular slip mechanism gives rise to the more realistic values of slip velocity compared to the prevailing basic surface hopping model.

## Materials and methods

An interesting non-Fickian self-diffusion mechanism of liquid molecules at an interface termed as bulk-mediated effective surface diffusion was proposed by Bychuk & O'Shaughnessy [15]. The mechanism consists of the repeated adsorption-desorption of a fluid molecule on the surface. The process begins with the adsorption of a near-surface molecule from the bulk liquid, followed by desorption after a certain waiting time. During this period of desorption, the molecule re-joins and undergoes the diffusion within the bulk liquid. Subsequently, the molecule is re-adsorbed at a different adsorption site when it lies within the attraction range of the substrate, after having travelled a certain distance in the bulk. The continuous cycle of adsorption and desorption effectively results in an interfacial self-diffusion process. This mechanism is unique in that the surface diffusion conforms to a Levy process instead of the usual Fick's law, exhibiting the superdiffusive behaviour with displacement $r \propto t$ instead of the familiar $r \propto \sqrt{t}$. Berezhkovskii *et al*. [16] found that bulk-mediated diffusion was most significant at a specific bulk layer thickness.

The bulk-mediated model above describes the self-diffusion of liquid molecules in a quiescent liquid. Here, we consider a flowing bulk liquid, or in other words, an external force that drives the desorbed molecule in the direction of the force when it returns to the bulk phase before being re-adsorbed. The bulk-mediated diffusion process in the presence of a driven flow is sketched in Fig 1. Intuitively, this should produce a faster molecular slip velocity compared to the surface hopping mechanism.

**Fig 1. Desorption mediated mechanism of molecular fluid slip: (a) adsorption/re-adsorption from bulk flow (b) adsorbed phase of duration $t_{surf}$ (c) desorption into bulk flow (d) bulk excursion of duration $t_{bulk}$ where $t_{total} = t_{surf} + t_{bulk}$**

Molecules lying within the surface attraction region of height $\lambda$ normal to the surface are adsorbed at a rate $Q_{ads}$. The characteristic time scale of re-adsorption $t_{ads}$ can be estimated from the displacement in the normal direction which occurs *via* diffusion

$$t_{ads} = \frac{D}{(Q_{ads}\lambda)^2},$$ (1)

where $D$ represents the bulk diffusivity.

From the survival probability $S(\tau) \propto \frac{1}{\sqrt{\tau}}$, the re-adsorption time distribution $\psi(\tau)$ can be expressed as

$$\psi(\tau) = -\frac{dS}{d\tau} = \frac{\sqrt{t_{ads}}}{\tau\sqrt{\tau}},$$ (2)

which is valid in the range $\tau > t_{ads}$. The total duration of time spent in the bulk liquid $t_{bulk}$ by an adsorbed molecule after the $n$ cycles of adsorption-desorption is

$$t_{bulk} = \int_{t_{ads}}^{t_{bulk}} n\,\tau\,\psi(\tau)\,d\tau = n^2 t_{ads}.$$ (3)

The total residence time $t_{surf}$ is

$$t_{surf} = \frac{n}{Q_{des}},$$ (4)

where $Q_{des}$ is the desorption rate. Hence, the total time $t_{total}$ during which the molecule undergoes bulk-mediated diffusion is

$$t_{total} = t_{surf} + t_{bulk}.$$ (5)

In general, the bulk-mediated diffusion dominates when $t_{total} \ll t_{ret}$, where $t_{ret}$ is termed as the surface retention time

$$t_{ret} = \frac{1}{Q_{des}^2 t_{ads}}.$$ (6)

In other words, surface diffusion *via* this mechanism takes place when the time between desorption events is longer than the re-adsorption time. This indicates a strongly adsorbing system in which the molecules are repeatedly re-adsorbed at a different adsorption site after getting desorbed without being permanently retained in the bulk phase. Molecules are lost to the bulk phase at times exceeding the surface retention time.

Following the above analysis, the molecular slip in bulk-mediated surface diffusion can be obtained as the total displacement of the adsorbed molecule in the direction parallel to the surface $n\langle x \rangle$ per unit time for the total duration of time spent in the bulk-mediated diffusion process

$$\langle v \rangle = \frac{n\langle x \rangle}{t_{total}}.$$ (7)

When a molecule is temporarily desorbed into the bulk flow, it is driven by the shear flow, which is assumed in this case to be linear. The driving force can be approximated by

$$F_{shear} = \mu A \dot{\gamma}, \tag{8}$$

where $\mu$ refers to the dynamic viscosity of the liquid, $A$ the effective molecular surface area and $\dot{\gamma}$ the shear rate of the liquid near the surface.

The net displacement of the molecule during each desorption/re-adsorption cycle can be estimated kinematically as

$$\langle x \rangle = \frac{1}{2m} F_{shear} t_{bulk}^2, \tag{9}$$

where $m$ refers to the mass of the molecule. Combining Eqs 3 to 9, the molecular slip velocity is eventually obtained as

$$\langle v \rangle = \frac{\mu A}{2m} \left( \frac{n t_{bulk}^2}{t_{surf} + t_{bulk}} \right) \dot{\gamma} \approx \frac{\mu A}{2m} n^3 t_{ads}^2 Q_{des} \dot{\gamma}. \tag{10}$$

The only unknown parameter in Eq 10 is $n$, the number of desorption/re-adsorption cycles.

# Results

In Figs 2 and 3, we compare the theoretical predictions of interfacial molecular slip from bulk-mediated diffusion in Eq 10 and surface hopping diffusion against two sets of experimental data from the literature for the slip velocity measurements of DI water in hydrophilic and hydrophobic microchannels conducted by Huang *et al.* [17]. Slip *via* surface hopping is given by the expression

$$u_m = u_h \sinh\left( \frac{\dot{\gamma}_s}{\dot{\gamma}_0} \right), \tag{11}$$

where the free surface diffusion velocity $u_h = v_0 a\, e^{-\frac{E_{a,m}}{k_B T}}$ and characteristic shear rate $\dot{\gamma}_0 = \frac{2 k_B T}{\mu A_{eff} a}$. The parameters used in Eq 10 are as follows: $\mu = 9 \times 10^{-4}$ kg/(s·m), $A = 1.617 \times 10^{-19}$ m² [10], $m = 3 \times 10^{-23}$ kg, $t_{ads} = 10^{-13}$ s [18], $Q_{des} t_{ads} = 10^{-6}$ [15]. Parameters for the surface hopping model are obtained from Yang [11].

**Fig 2. Slip velocity as a function of surface shear rate.** Solid line: theoretical prediction from desorption mediated diffusion mechanism for $n = 10^4$. Dashed line: theoretical prediction from surface hopping mechanism [11]. Symbols: experimental data for DI water in hydrophilic PDMS microchannel [17].

**Fig 3. Slip velocity as a function of surface shear rate.** Solid line: theoretical prediction from desorption mediated diffusion mechanism for $n = 9.5 \times 10^3$. Dashed line: theoretical prediction from surface hopping mechanism [11]. Symbols: experimental data for DI water in hydrophobic PDMS microchannel [17].

# Discussion

It can be observed from Figs 2 and 3 that the bulk-mediated surface diffusion mechanism of molecular slip is capable of producing much higher molecular slip velocities compared to the surface hopping diffusion model. However, this is subject to the number of 'bulk excursions' $n$ that the adsorbed molecule performs during the desorption/re-adsorption phase.

The theoretical prediction of molecular slip velocity exhibits a good match with the experimental data of Huang *et al.* [17] using a value of $n = 10^4$ and $n = 9.5 \times 10^3$ for the hydrophilic and hydrophobic microchannels respectively. The values of $n$ are consistent with the effect of surface wettability. For a hydrophilic surface, the stronger liquid-solid affinity should result in a higher number of re-adsorption events as opposed to a hydrophobic surface [19]. In contrast, the surface hopping diffusion model greatly under-estimates the slip velocity, indicating that this mode of molecular motion is less likely to occur on the liquid-solid pair studied in the experiments. The large disparity between the two models can be attributed to the higher drift velocity due to superdiffusive phenomenon in bulk-mediated surface diffusion.

Intriguingly, the stronger effect of bulk-mediated diffusion mechanism of slip on hydrophilic surfaces suggests that large slip velocities are possible on wetting surfaces since they fulfill the criteria of strong adsorbers due to their high affinity for liquid molecules. In this way, the slip velocity could be much higher than surface molecular motion *via* the surface hopping mechanism. The alteration of surface chemistry through artificial methods may also promote this mechanism of slip.

At increased surface shear rates, however, the deviation of the theoretical prediction from experimental data is palpable. The bulk-mediated surface diffusion model is linear in nature and therefore inadequate at the onset of non-linear slip behaviour. At present, there is no consensus on the non-linear dependence of slip velocity on shear rate observed experimentally although possible factors such as a less viscous layer of nanobubbles have been suggested [20].

The type of surface diffusion mechanism undergone by the adsorbed molecules is dictated by the liquid-substrate pair, depending on the nature of the fluid-substrate interactions, surface chemistry and relative time scales of both solid and fluid molecular motion. Bulk-mediated surface diffusion is only possible for fluid-substrate pairs possessing the characteristics of strong adsorbers. Surface retention times should typically be much higher than the desorption times so that each molecule spends more time on the surface than in the bulk and furthermore goes through a prolonged series of re-adsorption events without being instantly relinquished to the bulk phase upon desorption. Yet, if the surface binding energy is too high, desorption events become rarer and the bulk-mediated diffusion becomes ineffective.

In summary, a different mechanism of direct molecular motion on the substrate has been explored in this paper. The bulk-mediated diffusion model considers a sequence of periodic re-adsorption. This form of molecular slip motion displays higher slip velocities under specific conditions compared to the thermal surface diffusion model considered previously [21,22].

# Nomenclature

$A$      Effective molecular surface area

$D$      Bulk diffusivity

$F_{shear}$      Driving force

| | |
|---|---|
| $m$ | Molecular mass |
| $n$ | Cycles of adsorption-desorption |
| $Q_{ads}$ | Adsorption rate |
| $Q_{des}$ | Desorption rate |
| $r$ | Displacement |
| $S(\tau)$ | Survival probability |
| $t_{ads}$ | Characteristic time scale of re-adsorption |
| $t_{bulk}$ | Bulk liquid residence time |
| $t_{ret}$ | Surface retention time |
| $t_{surf}$ | Surface residence time |
| $t_{total}$ | Bulk-mediated diffusion time |
| $u_h$ | Free surface diffusion velocity |
| $u_m$ | Surface hopping slip velocity |
| $\dot{\gamma}$ | Shear rate |
| $\lambda$ | Normal distance from wall |
| $\mu$ | Dynamic viscosity |
| $v$ | Slip velocity |
| $\psi(\tau)$ | Re-adsorption time distribution |

# Author Contributions

**Conceptualization:** JJS JBMT.

**Methodology:** JJS JBMT.

**Software:** JBMT.

**Validation:** JJS JBMT.

**Formal analysis:** JJS JBMT.

**Investigation:** JJS JBMT.

**Resources:** JJS JBMT WKC.

**Data curation:** JBMT.

**Writing (original draft preparation):** JJS JBMT.

**Writing (review and editing):** JJS JBMT.

**Visualization:** JBMT.

**Supervision:** JJS WKC.

**Project administration:** JJS WKC.

**Funding acquisition:** JJS WKC.

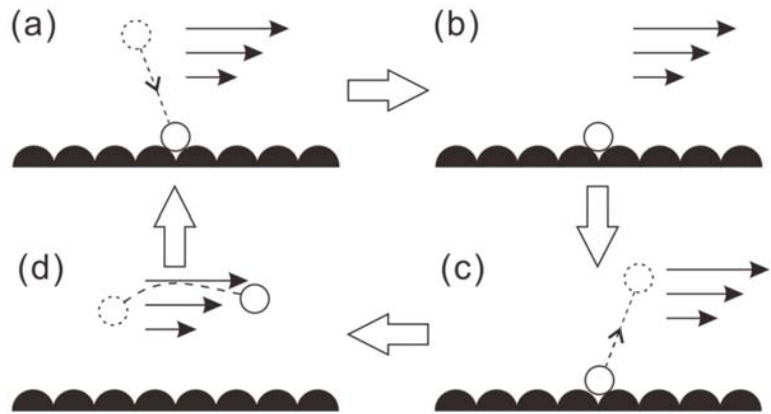

**Fig 1. Desorption mediated mechanism of molecular fluid slip: (a) adsorption/re-adsorption from bulk flow (b) adsorbed phase of duration $t_{surf}$ (c) desorption into bulk flow (d) bulk excursion of duration $t_{bulk}$ where $t_{total} = t_{surf} + t_{bulk}$**

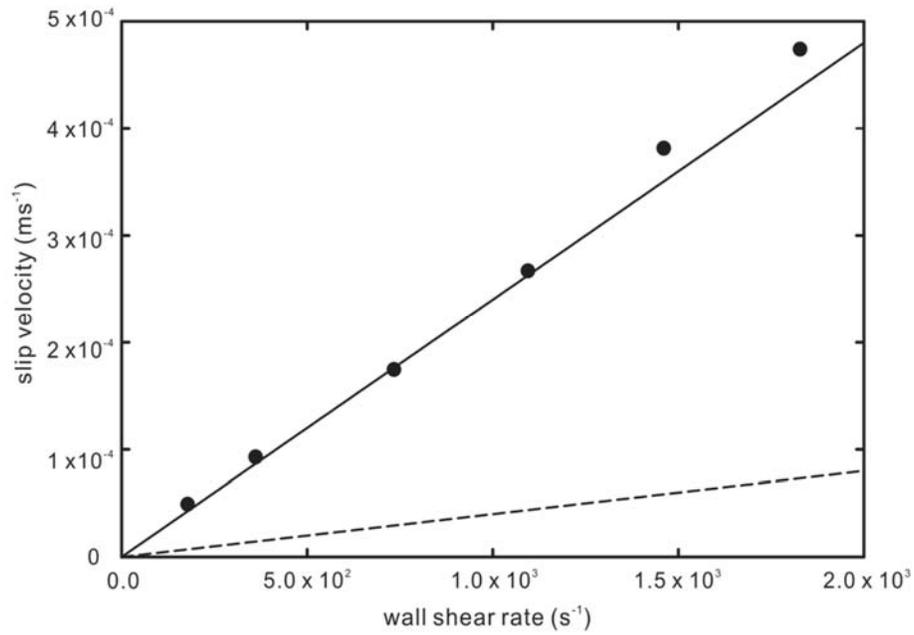

**Fig 2. Slip velocity as a function of surface shear rate.** Solid line: theoretical prediction from desorption mediated diffusion mechanism for $n = 10^4$. Dashed line: theoretical prediction from surface hopping mechanism [11]. Symbols: experimental data for DI water in hydrophilic PDMS microchannel [17].

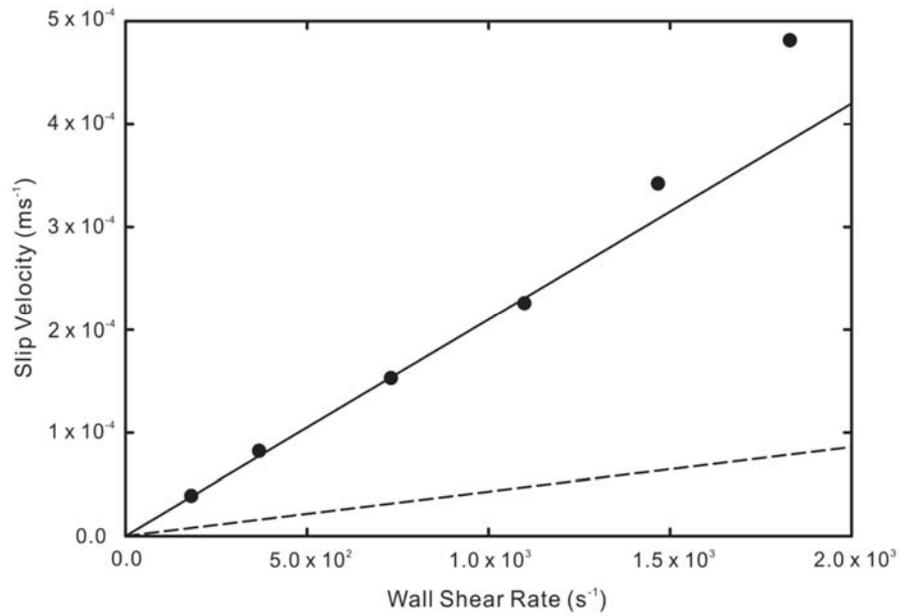

**Fig 3. Slip velocity as a function of surface shear rate.** Solid line: theoretical prediction from desorption mediated diffusion mechanism for $n = 9.5 \times 10^3$. Dashed line: theoretical prediction from surface hopping mechanism [11]. Symbols: experimental data for DI water in hydrophobic PDMS microchannel [17].